\documentclass[aip,apl,superscriptaddress,reprint]{revtex4-1}
\usepackage{graphicx}

\begin{document}

\title{Insulator-to-metal transition of SrTiO$_3$:Nb\\  single crystal surfaces induced by Ar$^+$ bombardment}

\author{C. Rodenb\"ucher}
\email[]{c.rodenbuecher@fz-juelich.de}
\affiliation{Forschungszentrum J\"ulich GmbH, Peter Gr\"unberg Institut 7, JARA, 52425 J\"ulich, Germany}
\author{S. Wicklein}
\affiliation{Forschungszentrum J\"ulich GmbH, Peter Gr\"unberg Institut 7, JARA, 52425 J\"ulich, Germany}
\author{R. Waser}
\affiliation{Forschungszentrum J\"ulich GmbH, Peter Gr\"unberg Institut 7, JARA, 52425 J\"ulich, Germany}
\affiliation{RWTH Aachen, Institut f\"ur Werkstoffe der Elektrotechnik 2, 52056 Aachen, Germany}
\author{K. Szot}
\affiliation{Forschungszentrum J\"ulich GmbH, Peter Gr\"unberg Institut 7, JARA, 52425 J\"ulich, Germany}
\affiliation{University of Silesia, A. Che\l kowski Institute of Physics, 40-007 Katowice, Poland}

%\date{\today}

\begin{abstract}
In this paper, the effect of Ar$^+$ bombardment of SrTiO$_3$:Nb surface layers is investigated on the macro- and nanoscale using surface-sensitive methods. After bombardment, the stoichiometry and electronic structure are changed distinctly leading to an insulator-to-metal transition related to the change of the Ti ``d'' electron from d$^0$ to d$^1$ and d$^2$. During bombardment, conducting islands are formed on the surface. The induced metallic state is not stable and can be reversed due to a redox process by external oxidation and even by self-reoxidation upon heating the sample to temperatures of 300$^\circ$C. \\
Copyright 2013 American Institute of Physics. This article may be downloaded for personal use only. Any other use requires prior permission of the author and the American Institute of Physics. The following article appeared in Appl. Phys. Lett. 102, 101603 and may be found at http://link.aip.org/link/?apl/102/101603.\end{abstract}

\maketitle

%\section{Introduction}
Perovskite oxides are very promising candidates for future redox-based resistive switching memories (ReRAM) due to the possibility of changing their resistance under external gradients related to an insulator-to-metal transition. In order to obtain insights into the basic physical mechanisms causing this transition, SrTiO$_3$ was investigated intensively thus serving as a reference material for perovskites. It was found that resistive switching due to electrical gradients is related to the movement of oxygen vacancies along extended defects \cite{Szot2006, Waser2007, Waser2009}. Local conductivity atomic force microscopy (LC-AFM) and surface spectroscopic studies provided valuable tools to elucidate the details of the resistive switching using SrTiO$_3$ single crystals \cite{Szot2006, Janousch2006} and thin films \cite{Szot2007, Muenstermann2008, Muenstermann2010}. Furthermore, the changes of resistivity upon application of external gradients were investigated intensively using Ar$^+$ bombardment and it was found that the conductivity of the SrTiO$_3$ surface can be dramatically increased due to changes of the chemical composition and oxygen stoichiometry \cite{Mukhopadhyay1993,Adachi1999,Albrecht2003,Kan2005,Reagor2005,Kan2007,Russell2008,Psiuk2009,Gross2011,Gross20112}. After this treatment, a conducting layer with the properties of a two-dimensional electron gas was detected \cite{Herranz2010,Bruno2011}. In this paper, we focus on the surface of Nb-doped SrTiO$_3$, which is of particular interest because of its electrical properties. Although Nb acts as a donor, which should lead to metallic conductivity, it was found that the surface layer becomes  highly resistive as soon as it comes into contact with oxygen\cite{Haruyama1997}. Hence, we investigated the influence of Ar$^+$ bombardment on the chemical composition and electronic structure by X-ray photoemission spectroscopy (XPS) as well as the transport properties on the macro- and nanoscale by electrical four-point measurements and LC-AFM.

%\section{Methods}
The measurements were conducted on epi-polished SrTiO$_3$:Nb with doping concentrations of  1.4 at\% purchased from Mateck. 
Ar$^+$ bombardment was performed using a Physical Electronics ion gun with an energy of 2~keV and an emission current of 15~mA. The beam was scanned over the sample in a raster of 4 x 4 mm. 
XPS spectra were recorded in situ by a Perkin Elmer instrument with monochromatized Al-K$_\alpha$ rays in ultrahigh vacuum (UHV) conditions. 
To investigate the temperature-dependent resistance before and after Ar$^+$ bombardment in situ, a special four-point measurement station in Valdes geometry was constructed. The four tips were aligned on a measurement head, which could be retracted during bombardment to avoid interference from electrode sputtering\cite{Psiuk2009}. By applying a voltage to the outer tips and measuring the potential between all four tips by electrometers it was possible to measure the absolute sheet resistance since the potential between the inner electrodes does not depend on the contact resistance.
Measurements of the topography on the nanoscale were performed using JEOL and Omicron atomic force microscopes (AFM)  with Pt/Ir tips under UHV conditions.
%\section{Results}
\begin{figure*}[ht]
\includegraphics[width=170mm]{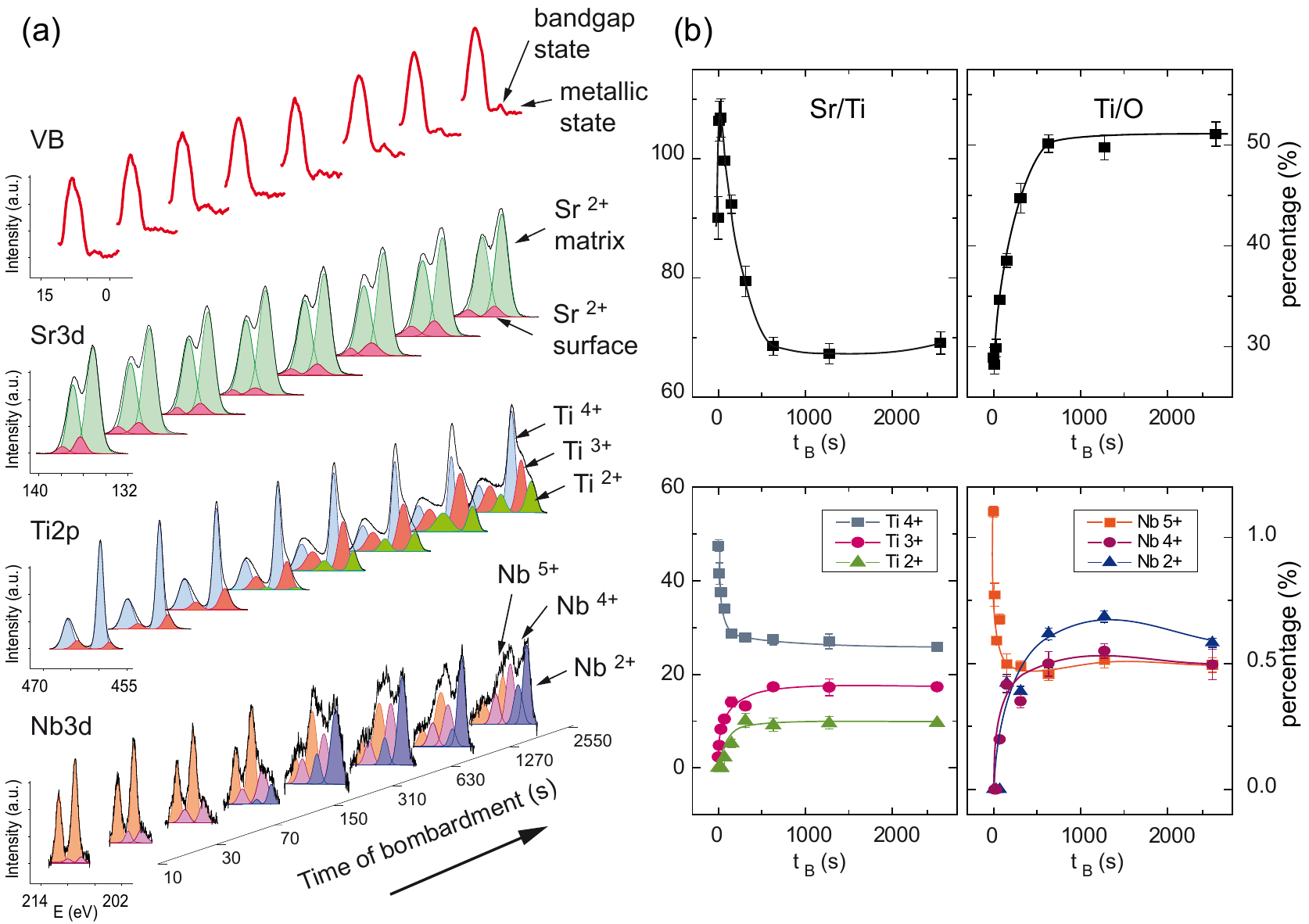}
\caption{(a) Electronic structure represented by the valence band and the core lines during Ar$^+$ bombardment. (b) Influence of the bombardment on the chemical composition and the amount of Ti and Nb valences measured by XPS.}
\label{xpsspec}
\end{figure*}
\begin{figure}[ht]
\includegraphics[width=85mm]{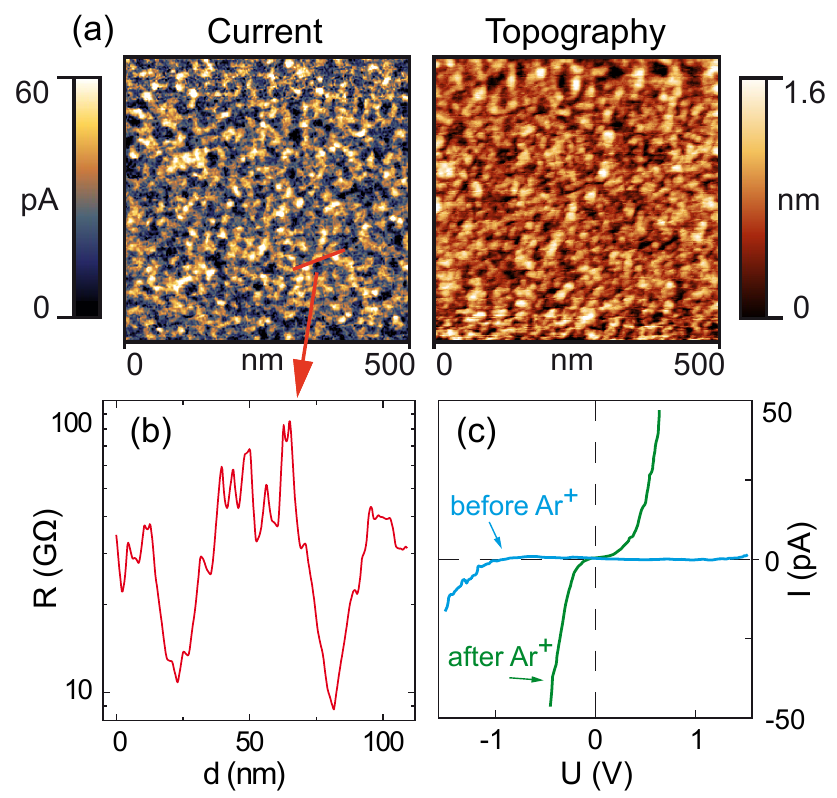}
\caption{LC-AFM measurements. (a) Local conductivity and topography after Ar$^+$ bombardment. (b) Local resistance along the red line in the AFM scan. (c) Comparison of IV-curves before and after Ar$^+$ bombardment.}
\label{afm}
\end{figure}
XPS measurements were conducted to investigate the changes in stoichiometry and electronic structure upon Ar$^+$ bombardment. After different bombardment times, the core lines of the spectrum were investigated as shown in Fig.~\ref{xpsspec}a. At the beginning, the as-received surface only displayed Ti with valence +4 and Nb with valence +5. Even after a short period of Ar$^+$ bombardment, the additional valences +3 and +2 for Ti and +4 and +2 for Nb were detected. Simultaneously, slight changes in the spectra of the bandgap close to the Fermi energy were observed, which could be an indication of the insulator-to-metal transition on the surface layer. Only the electronic structure of Sr was not significantly changed. It is in the valence state +2 and consists of two doublets. One large doublet (green) represents the Sr in the matrix of the crystal and one smaller doublet (red) reflects the change of the surroundings of the Sr atoms in the surface layer. Furthermore, the stoichiometry of the surface layer underwent a distinct modulation due to Ar$^+$ bombardment. The initial Sr excess was converted into a Ti excess and also the Ti/O ratio increased indicating that a preferential sputtering of Sr as well as of O occurs, which reduces the surface layer and alters the stoichiometry. After approximately 10 minutes (600~s) of bombardment the ratios of Sr/Ti and Ti/O became constant, which can be seen as a plateau in Fig.~\ref{xpsspec}b. This behavior correlates with the amount of additional valences of Ti and Nb obtained by peak fitting of the Ti 2p, Sr 3d and Nb 3d lines. The electrons of the Nb behave in a manner comparable to that of the Ti confirming that Nb substitutes the Ti and occupies the ``B'' side in the ABO$_3$ matrix \cite{Smyth1985}. Comparing the results obtained with the data of undoped SrTiO$_3$, it is striking that the insulator-to-metal transition induced by Ar$^+$ bombardment is very similar. This is quite surprising since Nb doping should already have induced metallicity. One possible explanation for this special role of the surface layer could be that the initial polishing of the crystals induces a high density of dislocations serving as electron traps. Another explanation could be that the surface layer with different properties than the bulk evolves due to the contact with oxygen as presented by Haruyama et al. \cite{Haruyama1997}.

Using LC-AFM, the changes of the topography and conductivity of the surface layer upon Ar$^+$ bombardment were investigated in situ (Fig.~\ref{afm}) to observe the nature of the insulator-to-metal transition on the nanoscale. During bombardment, the topography remained rather flat and unstructured, but a slight effect was observed regarding the RMS roughness parameter, which increased from 1.16~\AA~to 1.92~\AA. In contrast to the topography, the conductivity increased distinctly after 30 minutes of bombardment, which corresponds to the plateau region in the XPS measurement. The 2D map on the left side of Fig.~\ref{afm}a reveals that the conductivity is not homogeneous, but rather that islands of higher conductivity can be observed. This is illustrated by the extraction of the resistivity along a red line in the graphic. In Fig.~\ref{afm}b, it can be seen that the resistance changes locally by more than one order of magnitude within several tens of nanometers. The I-V characteristic obtained in the center of a conducting island (Fig.~\ref{afm}c) after bombardment shows a semiconducting characteristic, but the current is much higher than in the as-received case, which indicates that in some parts of the surface layer, probably corresponding to the Ti-rich phases, an insulator-to-metal transition started whereas other parts still remained in the semiconducting state. Furthermore, this assumption is supported by measurements of the crystallographic structure of the surface layer using low-energy electron diffraction (LEED). Before bombardment, a cubic diffraction pattern corresponding to the perovskite structure was visible but after bombardment no diffraction spots were observed which indicates that an amorphization of the surface layer has taken place. 

\begin{figure}[ht]
\includegraphics[width=85mm]{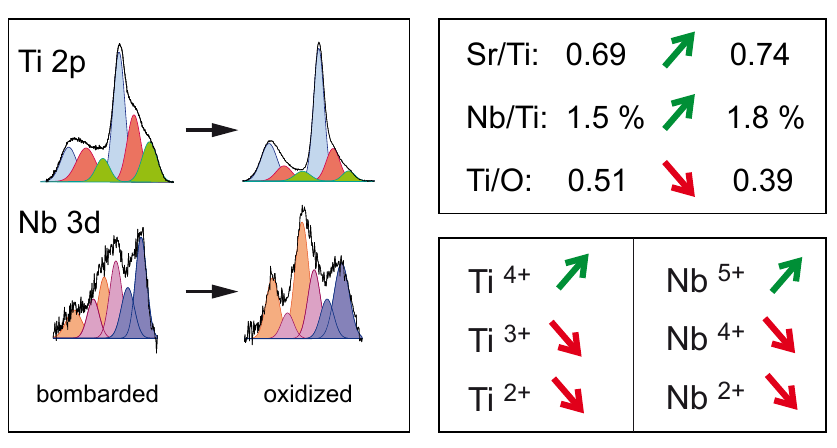}
\caption{Re-oxidation after Ar$^+$ bombardment by 10$^5$ L O$_2$ ($p_{O_2} = 1.3 \cdot10^{-3}$~mbar). Left: XPS measurement of the Ti and Nb corelines. Right: Development of the chemical composition and valences during oxidation.}
\label{oxy}
\end{figure}

In order to investigate the effect of re-oxidation on the surface bombarded by Ar$^+$,  we exposed the sample in situ to 10$^5$ L of O$_2$ ($p_{O_2} = 1.3 \cdot10^{-3}$~mbar) at room temperature. As shown in Fig.~\ref{oxy}, the additional valences of Nb and Ti were distinctly decreased and also the stoichiometry started to return to the as-received state. This behavior is consistent with the measurements obtained by Haruyama et al.\cite{Haruyama1997}, who reported changes in the electronic structure of the surface of crystals which were cleaved in situ and then exposed to oxygen. The XPS measurements presented here were conducted on a sample with a doping concentration of 1.4~\%, but also samples with doping concentrations of 0.1~\% and 10.1~\% were investigated and qualitatively displayed the same behavior. 
In conclusion, we demonstrated that the surface layer can be altered distinctly. This alteration does not only comprise a removal of the surface layer which is always present on the as-received crystal, but also the creation of a new Ti-rich surface layer with metallic conductivity.
\begin{figure}[ht]
\includegraphics[width=85mm]{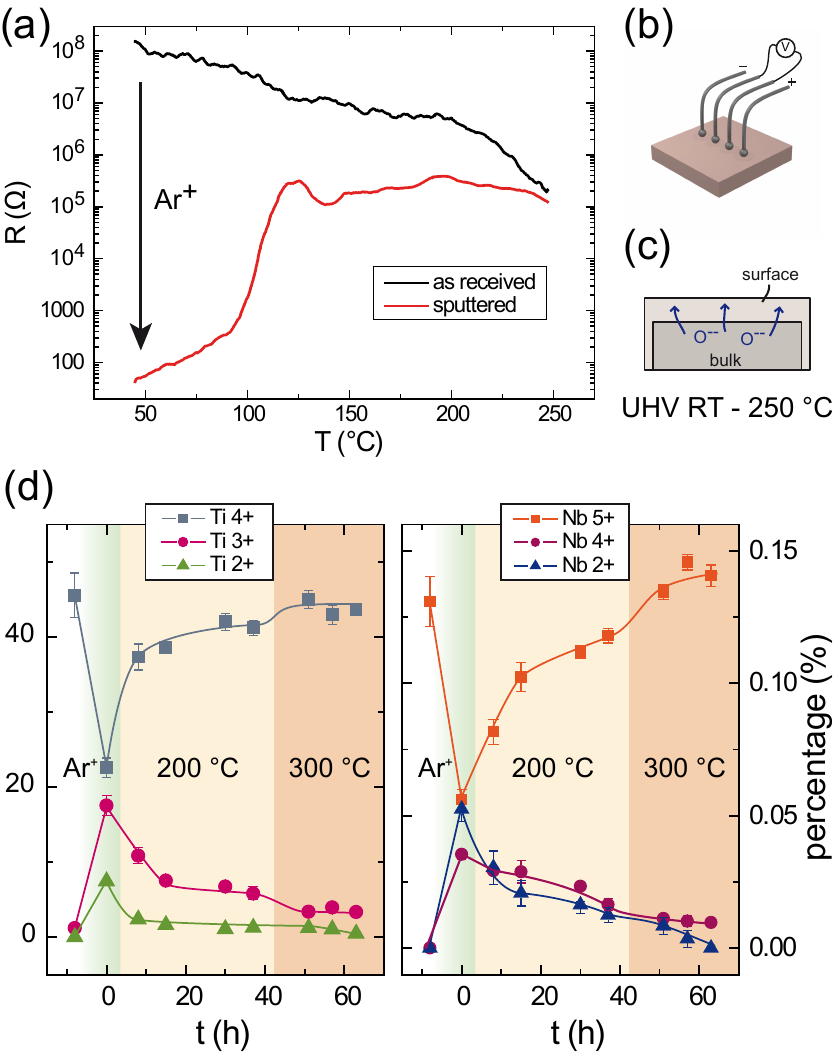}
\caption{(a) Temperature dependence of the resistance of the surface layer measured by the four-point method. (b) Overview of the alignment of the four retractable probes. (c) Illustration of the self-oxidation process. (d) Electronic structure measured by XPS as function of time and temperature showing a self-oxidation effect.}
\label{four}
\end{figure}

In order to investigate the metallic or non-metallic behavior of the surface layer resistance before and after bombardment, temperature-dependent four-point measurements were performed. As can be seen in Fig.~\ref{four}a, the as-received sample is semiconducting over the whole investigated temperature range proving that a surface layer exists on the nominally metallic material SrTiO$_3$:Nb. To start the insulator-to-metal transition, the surface was then bombarded with Ar$^+$ at room temperature for 30 minutes in situ. After this, the four contacts were lowered onto the sample surface as illustrated in Fig.~\ref{four}b and the four-point resistance was measured. During bombardment, the resistance dropped more than six orders of magnitude reflecting an insulator-to-metal transition. However, upon subsequent heating the resistance increased exponentially and above 100~$^\circ$C the resistance jumped to a value of more than 10$^5$~$\Omega$ and finally returned almost completely to the semiconducting state. Only if the sample is bombarded using a high-power ion gun with a beam current of 5 mA can a stable metallic state with a four-point resistance below 1~$\Omega$ be reached which does not show reoxidation effects.
The reoxidation of the intermediate state after mild bombardment can be understood by regarding the electronic structure measured by XPS. It can be seen in Fig.~\ref{four}d that immediately after bombardment additional valences for Nb and Ti are present.  Subsequently, the time dependence of the electronic structure was measured while the temperature was simultaneously increased. The additional valences created during the bombardment step vanish after a certain time and the electronic structure of the as-received sample is recovered. We assume that the bulk of the sample serves as a source of oxygen for the thermally induced self-reoxidation of the surface layer (cf. Fig.~\ref{four}). We also assume that this is a similar effect to that found as self-neutralization during photoemission experiments due to the movement of oxygen along dislocations\cite{Szade2009}.

%\section*{Conclusion}

In conclusion, we found that Ar$^+$ bombardment of SrTiO$_3$:Nb leads to a transformation of the as-received semiconducting surface layer into a metallic state related to the formation of Ti-rich phases. Investigations on the nanoscale revealed the generation of coexisting islands with different conductivities. Furthermore, it was shown that the transformation can be reversed by heating the sample leading to self-reoxidation of the surface layer. These measurements show that the surface of SrTiO$_3$:Nb can be altered easily by preferential sputter processes. This must be borne in mind when using the material as a substrate for the deposition of thin films.

%\begin{acknowledgments}
We thank J. Friedrich, M. Gerst and S. Masberg for construction of the four-point measurement station and technical support. 
This work was supported in part by the Deutsche Forschungsgemeinschaft (SFB 917).
%\end{acknowledgments}

\end{document}